\documentstyle[12pt]{article}
\begin{document}
\begin{center}
{\bf{STABLE CHROMOMAGNETIC QCD VACUUM AND CONFINEMENT}}

\vspace{1.5cm}

R.Parthasarathy{\footnote{e-mail address: sarathy@cmi.ac.in; sarathy@imsc.res.in}} \\
Chennai Mathematical Institute \\
H1, SIPCOT IT Park \\
Padur Post \\
Siruseri 603 103, India \\ 

\end{center}

\vspace{2.0cm}

{\noindent{\it{Abstract}}}

\vspace{0.5cm}

{\it{The stable chromomagnetic vacuum for $SU(2)$ Yang-Mills theory found earlier is shown to give 
a model for confinement in QCD,  using Wilson loop and a linear potential (in the leading 
order) for quark-antiquark interaction. The coefficent $k$ in this potential is found to be $\sim 
0.25\ GeV^2$, in satisfactory agreement with non-relativistic potential model calculations for 
charmonium. At finite temperature, the real effective energy density found earlier is used to 
obtain estimates of the deconfining temperature agreeing reasonably with lattice study for 
$SU(2)$}}.

\vspace{1.0cm} 

The economical definition of confinement of quarks in QCD is the 'area law' for the Wilson 
loop. The gauge invariant Wilson loop is 
\begin{eqnarray}
W(C)&=&Tr\ P\ e^{-ig\oint \ dx^{\mu}\ A_{\mu}^a(x)t^a},
\end{eqnarray}
where $P$ denotes the path ordering and $t^a$ are the generators of the gauge group. We shall 
consider $SU(2)$ Yang-Mills theory and choose the Savvidy [1] classical backgroound
\begin{eqnarray}
{\bar{A}}^a_0\ =\ 0 &;& {\bar{A}}^a_i\ =\ {\delta}^{a3}\ (-\frac{Hy}{2},\frac{Hx}{2},0).
\end{eqnarray}
This choice solves the classical equation of motion ${\bar{D}}^{ab}_{\mu}{\bar{F}}^{\mu\nu b}
=0$, where ${\bar{D}}^{ab}_{\mu}={\partial}_{\mu}{\delta}^{ab}+g{\epsilon}^{acb}{\bar{A}}
_{\mu}^c$. The classical background corresponds to constant chromomagnetic field in the 
third color direction ${\bar{F}}_{12}^3=H$ and this comes from the derivative terms in 
${\bar{F}}_{\mu\nu}^a={\partial}_{\mu}{\bar{A}}_{\nu}^a-{\partial}_{\nu}{\bar{A}}_{\mu}^a
+g{\epsilon}^{abc}{\bar{A}}_{\mu}^b{\bar{A}}_{\nu}^c$. For this reason, the Savvidy ansatz (2) 
is called 'Abelian like'. So the classical background (2) is esssentially Abelian-like, 
taking values in the Cartan subgroup of $SU(2)$.

\vspace{0.5cm}

The use of Abelian-like field strength can be understood from the idea of 't Hooft [2] who proposed 
'Abelian Projection'. This is a particular gauge fixing, breaking the gauge group $SU(N)$ to 
the maximal torus subgroup $H=U(1)^{N-1}$. For $SU(2)$, $H=U(1)$. This is realized in a 
specific gauge called the 'Maximal Abelian Gauge'. In the continuum formulation, this has the 
form 
\begin{eqnarray}
({\partial}_{\mu}{\delta}^{ab}+g{\epsilon}^{a3b}{\bar{A}}_{\mu}^3){\bar{A}}_{\mu}^b&=&0,
\end{eqnarray}
and the classical Savvidy background (2) satisfies (3). Numerical simulations on the lattice 
have found that the Abelian projected Wilson loop defined by $A_{\mu}^3$ exhibits the 
'area law' [3]. So (1) becomes 
\begin{eqnarray}
W(C)&=&\langle e^{-ig\oint dx^{\mu}\ A_{\mu}^3}\rangle, \nonumber \\
    &=&\langle e^{-i\frac{g}{2}\int_{S} dS^{\mu\nu}F_{\mu\nu}^3}\rangle,\nonumber \\
    &=&\langle e^{-i\frac{g}{2}H\times \ area}\rangle, 
\end{eqnarray}
where in (4), $H$ should correspond to the {\it{minimum value of the energy density}} as 
$W(C)$ involves vacuum expectation value. 

\vspace{0.5cm}

The classical energy density for the background (2), in the Euclidean formulation, is 
${\cal{E}}=\frac{H^2}{2}$. This energy density has a minimum ${\cal{E}}=0$ at $H=0$ and so 
$W(C)$ in (4) does not give the area law. {\it{In order to realize the area law from (4), the 
minimum energy density should correspond to $H\neq 0$.}} 

\vspace{0.5cm}

Savvidy [1] has studied the quantum 1-loop effective energy density which has a minimum 
lower than the above classical minimum and for which $H\neq 0$. However, Nielsen and Olesen [4]
pointed out that the 1-loop effective energy density in the background (2) had an imaginary 
part, stemming from the lowest Landau level and so the vacuum (ground state) of such a model 
is unstable. Various attempts were made to circumvent this sensitive issue which inhibited the 
progress. Instead of using the background (2), constant non-Abelian background ${\bar{A}}_0^a
=0$; ${\bar{A}}_i^a=K{\delta}_i^a$ was tried [5] and the said instability persisted. All 
these calculations were performed in the Gaussian (keeping only the terms quadratic in quantum 
fluctuations) approximation. 

\vspace{0.5cm}

We [6] have reexamined this important issue by retaining {\it{all}} the terms in the quantum 
fluctuations. Besides quadratic terms, there are terms cubic and quartic in quantum 
fluctuations. The detail of these calculations are given in Ref.6, {\it{in which the effective energy 
density has been shown to be real}}. Briefly, the Euclidean functional integral for $SU(2)$ pure Yang-Mills 
theory 
\begin{eqnarray}
Z&=&\int [dA_{\mu}^a]\ e^S, \nonumber \\
S&=&\int d^4x \{-\frac{1}{4}F_{\mu\nu}^aF_{\mu\nu}^a\}, \nonumber \\
F_{\mu\nu}^a&=&{\partial}_{\mu}A_{\nu}^a-{\partial}_{\nu}A_{\mu}^a+g{\epsilon}^{abc}A_{\mu}^b
A_{\nu}^c, 
\end{eqnarray}
is expanded around the classical background ${\bar{A}}_{\mu}^a$ in (2) as
\begin{eqnarray}
A_{\mu}^a&=&{\bar{A}}_{\mu}^a+a_{\mu}^a,
\end{eqnarray}
and the quantum fluctuations $a_{\mu}^a$ are taken to satisfy the 'background gauge' 
\begin{eqnarray}
{\bar{D}}_{\mu}^{ab}a_{\mu}^b&=&0.
\end{eqnarray}
This gauge choice is important. First of all, there is no Gribov ambiguity [7] in using this 
background gauge. It has been shown by Amati and Rouet [8] that the multiplicity of classical 
solutions satisfying the gauge condition is an irrelevant issue for quantizing non-Abelian 
Yang-Mills theories in the background gauge and an unambiguous generating functional is now 
possible. The correct treatment of the zero modes of the 1-loop operator gives the 
background gauge relative to the classical solution. See also [9]. The crucial point is that under a 
gauge transformation $U$, the quantum fluctuations $a_{\mu}^a$ in (6) transform 
homogeneously, namely, $a_{\mu}\equiv a_{\mu}^at^a; a_{\mu}\rightarrow Ua_{\mu}U^{-1}$ [9]. 
Second, with the background gauge (7), and using (3), we have ${\bar{D}}_{\mu}^{ab}(
{\bar{A}}_{\mu}^b+a_{\mu}^b)=0$ and so the 'Maximal Abelian Gauge' or Abelian projection is 
realized for the full gauge field ${\bar{A}}_{\mu}^a+a_{\mu}^a$. 

\vspace{0.5cm}

Now using (6) and (7) in (5), the unambiguous Euclidean generating functional $Z$ becomes,
\begin{eqnarray}
Z&=& \int [da_{\mu}^a]e^{S'},
\end{eqnarray}
with 
\begin{eqnarray}
S'&=&\int d^4x\{ -\frac{1}{4}{\bar{F}}_{\mu\nu}^a{\bar{F}}_{\mu\nu}^a+\frac{1}{2}a_{\mu}^a
{\Theta}_{\mu\nu}^{ac}a_{\nu}^c+g{\epsilon}^{acd}({\bar{D}}_{\nu}^{ae}a_{\mu}^e)a_{\mu}^c
a_{\nu}^d \nonumber \\
&-&\frac{g^2}{4}\Big( (a_{\mu}^aa_{\mu}^a)^2-a_{\mu}^aa_{\mu}^ca_{\nu}^aa_{\nu}^c\Big)\} - 
\ell og\ det (-{\bar{D}}_{\mu}^{ab}{\bar{D}}_{\mu}^{bc}), 
\end{eqnarray}
where 
\begin{eqnarray}
{\Theta}_{\mu\nu}^{ac}&=&({\bar{D}}_{\lambda}^{ab}{\bar{D}}_{\lambda}^{bc}){\delta}_{\mu\nu}
+2g{\epsilon}^{aec}{\bar{F}}_{\mu\nu}^e.
\end{eqnarray}
In arriving at (9), we have introduced the gauge fixing and the Faddeev-Popov ghost Lagrangian 
for the background gauge (7) and integrated the ghost fields, resulting in the last term in 
(9). The expansion in (9) is exact. The purpose of writing $S'$ in the form above is to isolate 
the stable and unstable modes of ${\Theta}_{\mu\nu}^{ac}$. 

\vspace{0.5cm}

For the Savvidy background, ${\Theta}_{44}^{ac}={\Theta}_{33}^{ac}={\bar{D}}_{\lambda}
^{ab}{\bar{D}}_{\lambda}^{bc}$, so that their contributions to ${\Gamma}$ cancel the ghost
contribution. Further the non-vanishing $\Theta$'s are ${\Theta}_{ij}^{ac}$ for $i,j=1,2$. Their 
eigenmodes and eigenvalues are:

\begin{eqnarray}
a_1^3\pm i a_2^3 &:& {k_1}^2+{k_2}^2+{k_3}^2+{k_4}^2\ \ (plane\ waves), \nonumber \\
(a_1^1+ia_2^1)-i(a_1^2+ia_2^2) &:& (2n+1)gH+2gH+k_3^2+k_4^2,\ (stable) \nonumber \\
(a_1^1-ia_2^1)+i(a_1^2-ia_2^2) &:& (2n+1)gH+2gH+k_3^2+k_4^2,\ (stable) \nonumber \\
(a_1^1+ia_2^1)+i(a_1^2+ia_2^2) &:& (2n+1)gH-2gH+k_3^2+k_4^2,\ (unstable) \nonumber \\
(a_1^1-ia_2^1)-i(a_1^2-ia_2^2) &:& (2n+1)gH-2gH+k_3^2+k_4^2,\ (unstable). \nonumber
\end{eqnarray}

The last two eigenvalues become negative when $n=0$ and for low momenta. These are called the 
'unstable modes'. As we encounter logarithm of the eigenvalues, in the quadratic approximation,  
negative eigenvalues make it imaginary and hence the effective energy density becomes
complex indicating vacuum instability. This in the Gaussian approximation.

\vspace{0.5cm}

The stable modes (the first two eigenvalues and the last two with $n\neq 0$) can be safely 
treated in the quadratic approximation. The contribution from the stable 
modes (see Ref.6 for details) are found to be 
\begin{eqnarray}
&\frac{10 g^2H^2}{96{\pi}^2}\ \{\ell og\Big(\frac{gH}{ {\mu}^2}\Big)+C\}, 
\end{eqnarray}
where $C$ is a real (infinite) constant and ${\mu}^2$ is a dimensionful constant introduced to 
render the argument of the logarithm dimensionless.

\vspace{0.5cm}

For the unstable modes, we [6] considered the full action in (9). The unstable modes involve 
the Lorentz indices $1$ and $2$ and the $SU(2)$ indices $1$ and $2$, because the classical 
background (2) is in the third color direction and so the cubic term in (9), namely, 
${\epsilon}^{acd}({\bar{D}}_{\nu}^{ae}a_{\mu}^e)a_{\mu}^ca_{\nu}^d$ vanishes. The quartic 
term in (9) for the unstable modes is found to be $\frac{1}{8}(|a_u|^2)^2$ where $a_u$ is 
the unstable mode. The functional integral $Z$ for the unstable modes is evaluated in Ref.6 
and from this the finite part of the unstable mode contribution to the energy density is 
found to be 
\begin{eqnarray}
&\frac{g^2H^2}{8{\pi}^2}\ell og\Big( \frac{gH}{ {\mu}^2}\Big)-\frac{g^2H^2}{4{\pi}^2}
\ell og I, 
\end{eqnarray}
where 
\begin{eqnarray}
I&=&\int dc'\ e^{-\{ {c'}^2({k_3'}^2+{k_4'}^2-1)+\frac{g^2}{256{\pi}^2}{c'}^4\}}.
\end{eqnarray}
The integral $I$ is convergent irrespective of whether ${k_3'}^2+{k_4'}^2$ is $<$ or $>$ 1. 
Further $I$ is real, finite and {\it{independent}} of $H$. Adding (12) to (11) and 
including the classical energy density, the effective energy density is found to be 
\begin{eqnarray}
{\cal{E}}&=&\frac{H^2}{2}+\frac{11g^2H^2}{48{\pi}^2}\{\ell og \Big(\frac{gH}{ {\mu}^2}\Big)
+C'\},
\end{eqnarray}
where $C'$ includes the second term in (12) along with $C$ in (11). 
The real constant $C'$ is then fixed by Coleman-Weinberg normalization $\frac{\partial}
{\partial H^2}{\cal{E}}|_{gH={\mu}^2}=\frac{1}{2}$ as $-\frac{1}{2}$. Thus the effective 
energy density for $SU(2)$ Yang-Mills theory in Savvidy background becomes 
\begin{eqnarray}
{\cal{E}}&=&\frac{H^2}{2}+\frac{11g^2H^2}{48{\pi}^2}\{\ell og \Big(\frac{gH}{ {\mu}^2}\Big)
-\frac{1}{2}\}.
\end{eqnarray}
This is real. The above result is  non-Abelian gauge theory effect. 

\vspace{0.5cm}

Quarks (fermions) can be added by minimally coupling them with the background (2) and 
functionally integrating $\psi$ and $\bar{\psi}$ in $Z$. The only change is the replacement of 
$11$ in (15) by $(11-N_f)$ for $N_f$ quark flavors. The prefactor $\frac{11-N_f}{48{\pi}^2}$ 
can be obtained from group theory considerations as well. Extending to $SU(3)$, this factor becomes 
$\frac{33-2N_f}{96{\pi}^2}$. 

\vspace{0.5cm}

In contrast to the classical energy density, the effective energy density in (15) has a minimum 
at non-zero $H$. From (15), we have 
\begin{eqnarray}
\frac{\partial {\cal{E}}}{\partial H^2}&=&\frac{1}{2}+\frac{11g^2}{48{\pi}^2}\ \ell og\Big(
\frac{gH}{ {\mu}^2}\Big), \nonumber \\
\frac{ {\partial}^2{\cal{E}}}{\partial (H^2)^2}&=&\frac{11g^2}{96{\pi}^2H^2}\ >\ 0.
\end{eqnarray}
The energy density has a minimum. When quarks are included, in order to have a minimum energy 
density, $N_f<11$ for $SU(2)$ or $N_f\leq 16$ for $SU(3)$. The minimum occurs when 
\begin{eqnarray}
H&=&\frac{ {\mu}^2}{g}\ e^{-\frac{24{\pi}^2}{11g^2}},
\end{eqnarray}
with $11$ appropriately replaced when quarks are included. {\it{Thus vacuum expectation value 
$H$ (corresponding to the minimum of ${\cal{E}}$) is {\underline{not zero}} which gives the 
Wilson loop the 'area law' and hence confinement}}.{\footnote{In the case of QED, as photons do not 
have self-interactions, the electrons alone contribute to the effective energy density in the 
constant magnetic field background. In this case, the effective energy density will be ${\cal{E}}
=\frac{H^2}{2}-A(e^2H^2)\{\ell og\Big(\frac{eH}{ {\mu}^2}\Big)-\frac{1}{2}\}$ where $A$ is a positive 
constant. This energy density has a {\it{maximum}} at non-zero value of $H$. Interpretting the 
maximum of the energy density to correspond to  
excited state, the area-law gives confinement of electrons by a linear effective potential, in a 
constant magnetic field when the electrons are in the excited state, akin to magnetic confinement of 
plasma state. }} 

\vspace{0.5cm}

The minimum energy density using (16) in (15) is 
\begin{eqnarray}
{\cal{E}}_{min}&=&-\frac{11g^2H^2}{96{\pi}^2},
\end{eqnarray}
which is lower than the classical minimum. This is the energy of the vacuum in the pure 
$SU(2)$ Yang-Mills theory. 

\vspace{0.5cm}

The result that the minimum of the energy density occurs when $H\neq 0$ (17) and 
${\bar{F}}_{12}^3=H$, imply that the vacuum expectation value $\langle {\bar{F}}_{12}^3
{\bar{F}}_{12}^3\rangle\neq 0$. This indicates that $\langle g^2F_{\mu\nu}^aF_{\mu\nu}^a
\rangle \neq 0$, the occurence of 'gluon condensate'. In this case 
\begin{eqnarray}
\langle g^2F_{\mu\nu}^aF_{\mu\nu}^a\rangle &=&2g^2H_{min}^2\ =\ 2{\mu}^4\ e^{-\frac{24
{\pi}^2}{11g^2}}.
\end{eqnarray}
Instead of using the strong coupling $g$ which runs, we use the result from the Charmonium 
decay analysis, $\langle g^2F_{\mu\nu}^aF_{\mu\nu}^a\rangle \ \sim \ 0.5GeV^4$. Then, 
$gH\sim 0.5 GeV^2$. With this estimate 
\begin{eqnarray}
W(C)&\sim & e^{\frac{gH}{2}\ area}\ =\ e^{\sigma\ area},
\end{eqnarray}
where $\sigma = \frac{gH}{2}=0.25 GeV^2$. It is well known that the 'area law' corresponds to 
a linear potential and in the leading order $V=\sigma R$ where $R$ is the separation of 
static quark and anti-quark.{\footnote{The Wilson loop in (4) involves the area in the 
$(x-y)$ plane for the Savvidy background (2). Following M.Baker, J.S.Ball, N.Brambilla,
G.M.Prosperi and F.Zachariasen, Phys.Rev. {\bf{D54}} (1996) 2829, the closed loop is defined 
by the quark (antiquark) trajectories ${\vec{z}}_1(t)({\vec{z}}_2(t))$ running from ${\vec{y}}_1
$ to ${\vec{x}}_1$ (${\vec{x}}_2$ to ${\vec{y}}_2$) as $t$ varies from $t_i$ to $t_f$. The 
quark (antiquark) trajectories are the world lines $C_1$ (and $C_2$) running from $t_i$ to 
$t_f$ ($t_f$ to $t_i$). The world lines $C_1$ and $C_2$ along with two straight lines at 
fixed time connecting ${\vec{y}}_1$ to ${\vec{y}}_2$ and ${\vec{x}}_1$ to ${\vec{x}}_2$ then 
make up the contour. Parameterising the quark (antiquark) trajectories as: $y_1(-\frac{R}{2},
-\frac{T}{2}); x_1(-\frac{R}{2},\frac{T}{2}); x_2(\frac{R}{2},\frac{T}{2}); y_2(\frac{R}{2},
-\frac{T}{2})$, where $t_i=-\frac{T}{2}, t_f=\frac{T}{2}$ and $R$ is the separation of the 
quark from antiquark, the area of the loop is $RT$. So, $W(C)=e^{-i\frac{gH}{2}RT}$. The 
potential is $V=\frac{i}{T}\ell og W(C)$ in the limit $T\rightarrow \infty$ and so $V=
\frac{gH}{2}R$. }} 

\vspace{0.5cm}

For a linear potential $V=kr$, the non-relativistic potential model calculations give  the 
$c{\bar{c}}$ bound states for $k=0.272 GeV^2$ which agrees with our estimate of 
$\sigma$ as $0.25 GeV^2$.

\vspace{0.5cm}

Thus, the stable vacuum in the chromomagnetic background is very much indicative of 
confinement, giving in the leading order the linear potential whose parameter $k$ is 
satisfactorily obtained.

\vspace{1.0cm}

Now we consider the $SU(2)$ Yang-Mills theory at finite temperature.  
In the studies of the Savvidy vacuum at finite temperature in the {\it{Gaussian 
Approximation}}, the effective energy density involved a temperature dependent imaginary part 
[10]. This inhibited the progress of examining the phase transition. 
We [11] have extended our zero-temperature studies (including the cubic and quartic terms) to finite 
temperature with chemical potential for the $SU(2)$ gauge bosons.

\vspace{0.5cm}

A chemical potential for massless non-Abelian bosons has been introduced in [12,13],
by observing that there are conserved color charges $Q^a=\int d^3x j_0^a;\ j_{\mu}^a=f^{abc}
A_{\nu}^bF_{\nu\mu}^c$. For $SU(2)$, one chooses $Q^3$. The grand canonical partition function 
will now have $\mu Q^3$ in the hamiltonian. This leads to the result of using $A_0^a=-i\mu 
{\delta}^{a3}$. This is not possible for Abelian gauge bosons. 

\vspace{0.5cm}

The role of chemical potential as a constant term in $A_0^a$ is similar to the use of 
Polyakov loop specified by a constant $A_0^a$ field in the third color direction. Now the 
Savvidy background becomes 
\begin{eqnarray}
{\bar{ {\cal{A}}}}_{\mu}^a&=&{\delta}^{a3}\{\frac{i\mu}{g},-\frac{Hy}{2},\frac{Hx}{2},0\},
\end{eqnarray}
which gives ${\bar{F}}_{12}^3=H$ and which solves the classical equation of motion. 
The background covariant derivative (Euclidean) now is 
\begin{eqnarray}
{\bar{D}}_{\lambda}^{ab}&=&{\partial}_{\lambda}{\delta}^{ab}+g{\epsilon}^{a3b}{\bar{A}}_{\lambda}
^3+\mu {\epsilon}^{a3b}v_{\lambda},
\end{eqnarray} 
where $v_{\lambda}=(1,0,0,0)$.

\vspace{0.5cm}

Once again we have isolated the unstable modes and treated them including the cubic and quartic 
terms in the fluctuations. The involved calculation (see Ref.11 for details) leads to the result that the 
effective energy density is real. The result for the effective energy density, including the zero temperature 
contribution is 
\begin{eqnarray}
{\cal{E}}&=&\frac{H^2}{2}+\frac{11(gH)^2}{48{\pi}^2}\Big(\ell og\Big(\frac{gH}{ {\Lambda}^2}\Big)
-\frac{1}{2}\Big)+\frac{ {\pi}^2}{45{\beta}^4} \nonumber \\
&+&\frac{(gH)^{\frac{3}{2}}}{\beta {\pi}^2}\sum_{\ell =1}^{\infty}\frac{\cos(\mu\beta\ell)}{\ell}
\{-\frac{\pi}{2}Y_1(\beta\ell\sqrt{gH})+K_1(\beta\ell\sqrt{gH}) \nonumber \\
&+&2\sum_{n=1}^{\infty}\sqrt{2n+1}K_1(\sqrt{2n+1}\beta\ell\sqrt{gH})\},
\end{eqnarray}
where $Y_1, K_1$ are modified Bessel functions. Setting $\beta=\frac{a}{\sqrt{gH}}$ and 
$\mu=b\sqrt{gH}$, the temperature dependence is written as 
\begin{eqnarray}
{\frac{ {\cal{E}}_T}{(gH)^2}}&=&\frac{ {\pi}^2}{45a^4}+\frac{1}{ {\pi}^2a}\sum_{\ell =1}^{\infty}
\frac{\cos(ab\ell)}{\ell}\{-\frac{\pi}{2}Y_1(a\ell)+K_1(a\ell) \nonumber \\
&+&2\sum_{n=1}^{\infty}\sqrt{2n+1}K_1(a\ell \sqrt{2n+1})\}.
\end{eqnarray}
 
\vspace{0.5cm}

In [11], we have plotted $\frac{ {\cal{E}}_T}{(gH)^2}$ with $T$ in units of $\sqrt{gH}$. The parameter $b$ 
involves the chemical potential. For $b=0$, zero chemical potential, the variation is smooth.
At high temperatures, the behaviour is like that of non-interacting relativistic gas.
For $b=1, 2, 3$ the variation shows a minimum and then smooth rise. A non-zero chemical 
potential thus triggers deconfinement phase transition. 

\vspace{0.5cm}

Deconfinement occurs for $b=1$ around $T/{\sqrt{gH}}\sim 0.4$ and for $b=2$ around 
$T/{\sqrt{gH}}\sim 0.7$. We have earlier identified the string tension $\sigma$ as $\frac{gH}{2}$ and 
so our results give for the deconfining temperature $\frac{T}{\sqrt{\sigma}}\sim 0.5656$ for $b=1$ and 
$0.9899$ for $b=2$. It is interesting to compare with the lattice studies. For $SU(2)$, Lucini, Teper and 
Wenger [14] find the deconfining temperature 
$\frac{T}{\sqrt{\sigma}}\sim 0.709$; the agreement is satisfactory for 
$1>b<2$. As the chemical potential $\mu=b\sqrt{gH}$, using $gH=0.5\ GeV^2$, the lowest value 
for the chemical potential triggering deconfinement is $0.7\ GeV<\mu<1.41\ GeV$.

\vspace{1.5cm}

To summarize, the stable chromomagnetic vacuum for $SU(2)$ Yang-Mills theory found in [6] gives 
a model for confinement using Wilson loop and hence a linear potential (in the leading order) 
for the quark-antiquark interaction. The coefficient $k$ in this potential is $\sim 0.25\ GeV^2$,
in satisfactory agreement with non-relativistic potential model for charmonium. At finite 
temperature, the real effective energy density found in [11] is used to obtain estimates of the 
deconfining temperature and this reasonably agrees with the lattice study for $SU(2)$. 

\vspace{0.5cm}

{\noindent{\bf{Acknowledgement}}}

\vspace{0.5cm}

The author thanks G.'t Hooft for very useful discussion when he visited Chennai Mathematical 
Institute during November 2009. The contents of this paper were presented by the author in his 
talk at the conference on ``Strong Interactions in the 21st Century'', held at the Tata 
Institute of Fundamental Research, Mumbai, during Feb. 10 - 12, 2010. The author thanks the 
organisers of the conference for warm hospitality and the Chennai Mathematical Institute for 
encouragement and travel support. 

\vspace{1.0cm} 

{\noindent{\bf{References}}}

\vspace{0.5cm}

\begin{enumerate}
\item G.K.Savvidy, Phys.Lett. {\bf{71B}} (1977) 133. 
\item G.'t Hooft, Nucl.Phys. {\bf{B190}} [FS3] (1981) 455. 
\item T. Suzuki and I. Yotsuyanagi, Phys.Rev. {\bf{D42}} (1990) 4257; \\
      K-I. Kondo and A. Shibata, hep-th/0801.4203.
\item N.K. Nielsen and P. Olesen, Nucl.Phys. {\bf{B144}} (1978) 376. 
\item R. Anishetty, Phys.Lett. {\bf{B108}} (1982) 295; \\
      R. Parthasarathy, M. Singer and K.S. Viswanathan, Can.J.Phys. {\bf{61}} (1983) 1442; \\
      S. Huang and A.R. Levi, Phys.Rev. {\bf{D49}} (1994) 6849. 
\item D. Kay, A. Kumar and R. Parthasarathy, Mod.Phys.Lett. {\bf{A20}} (2005) 1655; \\
      R. Parthasarathy and A. Kumar, Phys.Rev. {\bf{D75}} (2007) 085007. 
\item V.N. Gribov, Nucl.Phys. {\bf{B139}} (1978) 1. 
\item D. Amati and A. Rouet, Phys.Lett. {\bf{73B}} (1978) 39. 
\item R. Parthasarathy, {\it{Pramana}}. {\bf{32}} (1989) 563.
\item J.J. Kapusta, Nucl.Phys. {\bf{B190}} (1981) 425. \\
      B. Nuller and J. Rafelski, Phys.Lett. {\bf{101B}} (1981) 111. \\
      J. Chakrabarti, Phys.Rev. {\bf{D24}} (1981) 2232. \\
      M. Reuter and W. Dittrich, Phys.Lett. {\bf{144B}} (1984) 99. \\
      M. Ninomiya and N. Sakai, Nucl.Phys. {\bf{B190}} (1981) 316. \\
      A. Cabo, O.K. Kalashnikov and A.E. Shabad, Nucl.Phys. {\bf{B185}} (1981) 473. \\
      A.O. Starinets, A.S. Vshivtsev and V.C. Zhukovsky, Phys.Lett. {\bf{B322}} (1994) 403. \\
      P.N. Meisinger and M.C. Ogilvie, Phys.Rev. {\bf{D66}} (2002) 105006. 
\item R.Parthasarathy and A.Kumar, Phys.Rev. {\bf{D75}} (2007) 085007.
\item R. Anishetty, J.Phys. {\bf{G10}} (1984) 423. 
\item A. Actor, Phys.Lett. {\bf{157B}} (1985) 53. 
\item Lucini, M. Teper and Wenger, arXiv: hep-lat/0307017; 0502003.
\end{enumerate} 

\end{document}